\documentclass[cits]{PoS}

\title{Critical behaviour of the compact $3d$ $U(1)$ gauge theory at finite 
temperature}

\ShortTitle{Critical behaviour of the compact $3d U(1)$ gauge theory at finite 
temperature}

\author{Oleg Borisenko\\
        Bogolyubov Institute for Theoretical Physics, 
        National Academy of Sciences of Ukraine \\
        03680 Kiev, Ukraine \\
        E-mail: \email{oleg@bitp.kiev.ua}}

\author{\speaker{Roberto Fiore}\\
        Dipartimento di Fisica, Universit\`a della Calabria,
        and INFN - Gruppo Collegato di Cosenza \\I-87036 Rende, Italy\\
        E-mail: \email{fiore@cs.infn.it}}

\author{Mario Gravina\\
        Laboratoire de Physique Th\'eorique, 
        Universit\'e de Paris-Sud 11, B\^atiment 210 \\
        91405 Orsay Cedex, France\\
        E-mail: \email{Mario.Gravina@th.u-psud.fr}}

\author{Alessandro Papa\\
        Dipartimento di Fisica, Universit\`a della Calabria, 
        and INFN - Gruppo Collegato di Cosenza \\I-87036 Rende, Italy\\
        E-mail: \email{papa@cs.infn.it}}

\abstract{Critical properties of the compact three-dimensional $U(1)$ lattice 
gauge theory are explored at finite temperatures. The critical point of the 
deconfinement phase transition, critical indices and the string tension are 
studied numerically on lattices with temporal extension $N_t = 8$ and spatial 
extension ranging from $L = 32$ to $L = 256$. The critical indices, which 
govern the behaviour across the deconfinement phase transition, are generally 
expected to coincide with the critical indices of the two-dimensional $XY$ 
model. 
It is found that the determination of the infinite volume critical point 
differs from the pseudo-critical coupling at $L = 32$, found earlier in the 
literature and implicitly assumed as the onset value of the deconfined phase. 
The critical index $\nu$ computed from the scaling of the pseudocritical 
couplings agrees well with the value $\nu = 1/2$ of the $XY$ model. The 
computation of the index $\eta$ brings to a value larger than expected. The 
possible reasons for such behaviour are discussed.}

\FullConference{The XXVIII International Symposium on Lattice Field Theory, 
Lattice2010\\June 14-19, 2010\\Villasimius, Italy}

\begin{document}

\section{Introduction}

This article deals with the compact three-dimensional ($3d$) $U(1)$ lattice 
gauge theory (LGT), whose partition function can be written as
\begin{equation}
Z(\beta_t,\beta_s) = \int_0^{2\pi}\prod_{x\in\Lambda}\: \prod_{n=0}^2
\frac{d\omega_n (x)}{2\pi} \ \exp{S[\omega]} \ ,
\label{PTdef}
\end{equation}
where $\Lambda$ is an $L^2\times N_t$ lattice, $S$ is the Wilson action, 
which reads
\begin{equation}
 S[\omega] = \beta_s\sum_{p_s} \cos\omega (p_s) + 
\beta_t\sum_{p_t} \cos\omega (p_t) 
\label{wilsonaction} 
\end{equation}
and sums run over all space-like ($p_s$) and time-like ($p_t$) plaquettes.
The plaquette angles $\omega(p)$ are defined in the standard way. The 
anisotropic couplings $\beta_t$ and $\beta_s$ are defined in 
Ref.~\cite{beta_szero}.  To study the theory at finite temperature, periodic 
boundary conditions in the temporal direction are imposed on the gauge fields.

At zero temperature the theory is confining at all values of the bare coupling 
constant~\cite{polyakov}, while at finite temperature the theory undergoes a 
deconfinement phase transition. It is well known that the partition function 
of the $3d$ $U(1)$ LGT in the Villain formulation coincides with that of the 
$2d$ $XY$ model in the leading order of the high-temperature 
expansion~\cite{parga}. When combined with the universality conjecture by 
Svetitsky-Yaffe~\cite{svetitsky}, this result leads to conclude that the 
deconfinement phase transition belongs to the universality class 
of the $2d$ $XY$ model, which is known to have Berezinskii-Kosterlitz-Thouless 
(BKT) phase transition of infinite 
order~\cite{Berezinsky:1970fr,Kosterlitz:1973xp}. 
In particular, one might expect the critical behaviour of the Polyakov loop 
correlation function $\Gamma (R)$ to be governed by the following expressions
\begin{equation}
\Gamma (R) \ \asymp \ \frac{1}{R^{\eta (T)}} \ ,
\label{PLhight}
\end{equation}
for $\beta \geq \beta_c$ and 
\begin{equation}
\Gamma (R) \ \asymp \ \exp \left [ -R/\xi (t)  \right ] \ ,
\label{PLlowt}
\end{equation}
for $\beta < \beta_c$, $t=\beta_c/\beta -1$.
Here, $R\gg 1$ is the distance between test charges, $T$ is the temperature
and $\xi \sim e^{bt^{-\nu}}$ is the correlation length. Such behaviour of 
$\xi$ defines the so-called {\it essential scaling}. The critical indices 
$\eta (T)$ and $\nu$ are known from the renormalization-group (RG) analysis 
of the $XY$ model: $\eta (T_c) =1/4$ and $\nu=1/2$, where $T_c$ is the BKT 
critical point. 

The direct numerical check of these predictions was performed on lattices 
$L^2\times N_t$ with $L=16, 32$ and $N_t=4,6,8$ in Ref.~\cite{mcfinitet}. 
Though the authors of Ref.~\cite{mcfinitet} confirm the expected BKT nature 
of the phase transition, the reported critical index is almost three times 
that predicted for the $XY$ model, $\eta (T_c) \approx 0.78$. 
More recent numerical simulations of Ref.~\cite{chernodub} have been mostly 
concentrated on the study of the properties of the high-temperature phase. 
In these papers it was found that, for the isotropic lattice 
$\beta_s=\beta_t=\beta$ with $L=32$ and $N_t=8$, the pseudo-critical point 
is $\beta_{pc}=2.30(2)$ for Ref.~\cite{mcfinitet} and 
$\beta_c\approx 2.346(2)$ for Ref.~\cite{chernodub}. Values of $\beta$ above 
these values were taken implicitly as belonging to the deconfined phase.

In Ref.~\cite{beta_szero} we have studied the model on extremely anisotropic 
lattice with $\beta_s=0$. By a simple analytical analysis we showed that in 
the limits of both small and large $\beta_t$ such anisotropic model reduces to 
the $2d$ $XY$ model with some effective couplings. Then we performed numerical 
simulations of the effective spin model for the Polyakov loop which can be 
exactly computed in the limit $\beta_s=0$. We used lattices with $N_t=1,4,8$ 
and with the spatial extent $L\in [64,256]$ and found that the index $\eta$ 
is well compatible with the $XY$ value. We may thus assume that, at least in 
the limit $\beta_s=0$, the $3d$ $U(1)$ LGT does belong to the universality 
class of the $XY$ model. 

Here we consider the isotropic model on the lattice with $N_t=8$. Our strategy 
is the following: we postulate that the scaling laws of the $XY$ model are
valid and use them to determine the critical indices of the gauge model. 
In doing so we have encountered certain surprises: (i) the infinite volume 
critical coupling turned out to be essentially higher than the values for the 
pseudo-critical couplings reported in Refs.~\cite{mcfinitet,chernodub}. As a 
consequence, the values of $\beta$ used in Ref.~\cite{chernodub} to study the 
deconfinement phase lie well inside the confinement phase when the 
thermodynamic limit is considered; (ii) the index $\nu$ extracted from the 
scaling of the pseudo-critical couplings with $L$ does agree well with the 
expected $XY$ value $\nu=1/2$, but the index $\eta$ was found to be strikingly 
different from the $XY$ value, namely $\eta \approx 0.50$. While the value 
$\eta \approx 0.78$ obtained in Ref.~\cite{mcfinitet} could, in principle be 
attributed to rather small lattices used, $L=32$, and to an incorrect location 
of the critical point, our result is almost insensitive to varying the spatial 
extent if $L$ is large enough.  

\section{Numerical results}
\label{setup}

We simulated the system on lattices of the type $L^2 \times N_t$, with 
$N_t=8$ fixed and $L$ increasing towards the thermodynamic limit
(for details, see Ref.~\cite{isotropic}). In the adopted Monte Carlo algorithm 
a sweep consisted in a mixture of one Metropolis update and five microcanonical
steps. Measurements were taken every 10 sweeps in order to reduce the 
autocorrelation and the typical statistics per run was about 100k. The error 
analysis was performed by the jackknife method over bins at different blocking 
levels.

The observable used as a probe of the two phases of the finite temperature 
$3d$ $U(1)$ LGT is the Polyakov loop, defined as
\begin{equation}
P(\vec{x})=\prod_t U_0(\vec{x},t) \;,
\end{equation}
where $U_0(\vec{x},t)$ is the temporal link attached at the spatial point 
$\vec{x}$.
The effective theory for the Polyakov loop is two-dimensional and possesses 
global $U(1)$ symmetry. Since the global symmetry cannot be broken 
spontaneously in two dimensions owing to the Mermin-Wagner-Coleman theorem, 
the expectation value of the Polyakov loop vanishes in the thermodynamic 
limit. On a finite lattice $\langle \sum_{\vec{x}} P(\vec{x}) \rangle =0$ due 
to $U(1)$ symmetry (if the boundary conditions used preserve the symmetry). 
This is confirmed by the numerical analysis on the periodic lattice: 
in the confined (small $\beta$) phase the values taken by the Polyakov loop 
in a typical Monte Carlo ensemble scatter around the origin of the complex 
plane forming a uniform cloud, whereas in the deconfined (high $\beta$) phase 
they distribute on a ring, the thermal average being equal to zero in both 
cases. What really feels the transition is then the absolute value of $P$, 
which has been chosen to be the order parameter in this work. 

At finite volume the transition manifests through a peak in the magnetic 
susceptibility of the Polyakov loop, defined as
\begin{equation}
\chi_L=L^2 (\langle |P|^2  \rangle -\langle |P| \rangle^2) \quad , \quad 
P = \frac{1}{L^2} \ \sum_x P(\vec{x}) \quad .
\end{equation}
The value of the coupling at which this happens is the pseudo-critical 
coupling, $\beta_{pc}$. By increasing the spatial volume, the position of the 
peak moves towards the (nonuniversal) infinite volume critical coupling, 
$\beta_c$. The value of $\beta_{pc}$ for a given $L$ is determined by 
interpolating the values of the susceptibility $\chi_L$ around the peak by a 
Lorentzian function. In Table~\ref{beta_pc_tab} we summarize the resulting 
values of $\beta_{pc}$ and the peak values of the susceptibility $\chi_L$ for 
the several volumes considered in this work (we included also the 
determination for $L=32$, taken from the first paper in Ref.~\cite{chernodub}).

\begin{table}[ht]
\centering\caption[]{$\beta_{pc}$ and peak value of the Polyakov loop susceptibility 
$\chi_L$ on the lattices $L^2 \times 8$.}
\vspace{0.2cm}
\begin{tabular}{|c|l|c|}
\hline
 $L$ & $\beta_{pc}$ & $\chi_{L,\mbox{\scriptsize max}}$ \\
\hline
 32 & 2.346(2), Ref.~\cite{chernodub} & \\
 48 & 2.4238(67) & 12.93(41) \\
 64 & 2.4719(39) & 20.09(66) \\
 96 & 2.5648(96) & 38.8(1.6) \\
128 & 2.6526(59) & 60.1(3.5) \\
150 & 2.68(1)    & 92.6(8.0) \\
200 & 2.7336(69) & 144(12)   \\
256 & 2.7780(40) & 220(20)   \\
\hline
\end{tabular}
\label{beta_pc_tab}
\end{table}

There are, in principle, two hypotheses to be tested in order to locate 
the infinite volume critical coupling $\beta_c$: first order and BKT 
transition. The hypothesis of first order transition is not incompatible with 
data for the peak susceptibility for $L\geq 128$. However, the corresponding 
scaling law for the pseudo-critical couplings,
\begin{equation}
\beta_{pc}=\beta_c+\frac{A}{L^2} \quad ,
\label{b_pc_first}
\end{equation}
seems to be ruled out by our data ($\chi^2$/d.o.f equal to 5.6 for $L\geq 96$, 
3.7 for $L\geq 128$, 2.1 for $L\geq 96$).

Assuming the essential scaling of the BKT transition, {\it i.e.} 
$\xi \sim e^{bt^{-\nu}}$, the scaling law for $\beta_{pc}$ becomes
\begin{equation}
\beta_{pc}=\beta_c+\frac{A}{(\ln L + B)^{\frac{1}{\nu}}} \quad .
\label{b_pc}
\end{equation}

A 4-parameter fit of the data for $\beta_{pc}(L)$ given in 
Table~\ref{beta_pc_tab} with the law given in Eq.~(\ref{b_pc}) leads to 
unstable values of the parameters. Instead, when the parameter $\nu$ is fixed
at the $XY$ value, $\nu= 1/2$, the fit is stable for lattices with size
not smaller than $L$, leading to an estimated value of the infinite volume
critical coupling, $\beta_c=3.06(11)$ (see Ref.~\cite{isotropic}). 

By finite size scaling (FSS) analysis at $\beta_c$, we can extract other 
critical indices. An interesting one is the magnetic critical index, $\eta$, 
which enters the scaling law 
\begin{equation}
\chi_L(\beta_c) \sim L^{2-\eta} \quad .
\label{eta_scale}
\end{equation}
Actually in this law one should consider logarithmic corrections 
(see Refs.~\cite{Kenna-Irving,Hasenbusch} and references therein) and, indeed,
recent works on the $XY$ universality class generally include them. However,
taking these corrections into account for extracting critical indices calls
for very large lattices even in the $XY$ model; for the theory under 
consideration to be computationally tractable, we have no choice but to neglect
logarithmic corrections.

Setting the coupling $\beta$ at the value of our best estimation for $\beta_c$,
i.e. $\beta=3.06$, we determined the susceptibilities $\chi_L(\beta_c)$ for 
several volumes. Then, following FSS, we fitted the results with the law 
$\chi_L(\beta_c)=A L^{2-\eta}$ and got
\begin{equation} 
A=0.0171(10) , \quad \eta=0.496(15) \quad (\chi^2/\mbox{d.o.f.}=0.60) \quad .
\end{equation} 

This value for $\eta$ is by far incompatible with the 2$d$ $XY$ value, 
$\eta_{XY}=0.25$. The most extreme consequence of this finding is that the 
deconfinement transition in the 3$d$ $U(1)$ LGT at finite temperature does 
not belong to the same universality class as 2$d$ $XY$ spin model. 
This would contradict the Svetitsky-Yaffe conjecture, raising a problem in the 
understanding of the deconfinement mechanism in gauge theories. 
We will further comment on this issue in the discussion section.

In such a situation, it becomes particularly useful to have another 
determination of the index $\eta$, by an independent approach. 
Following Ref.~\cite{beta_szero}, we define an {\em effective} $\eta$ index, 
through the 2-point correlator of Polyakov loops, according to
\begin{equation}
\eta_{\mbox{\scriptsize eff}}(R) \equiv \frac{\log [\Gamma (R)/\Gamma (R_0)]}
{\log [R_0/R]} \quad ,
\label{eff_eta_def}
\end{equation}
with $R_0$ chosen equal to 10, as in Ref.~\cite{beta_szero}. This quantity is 
constructed in such a way that it exhibits a {\it plateau} in $R$ if the 
correlator obeys the law~(\ref{PLhight}), valid in the deconfined phase. 

\begin{figure}[tb]
\centering
\includegraphics[width=15cm]{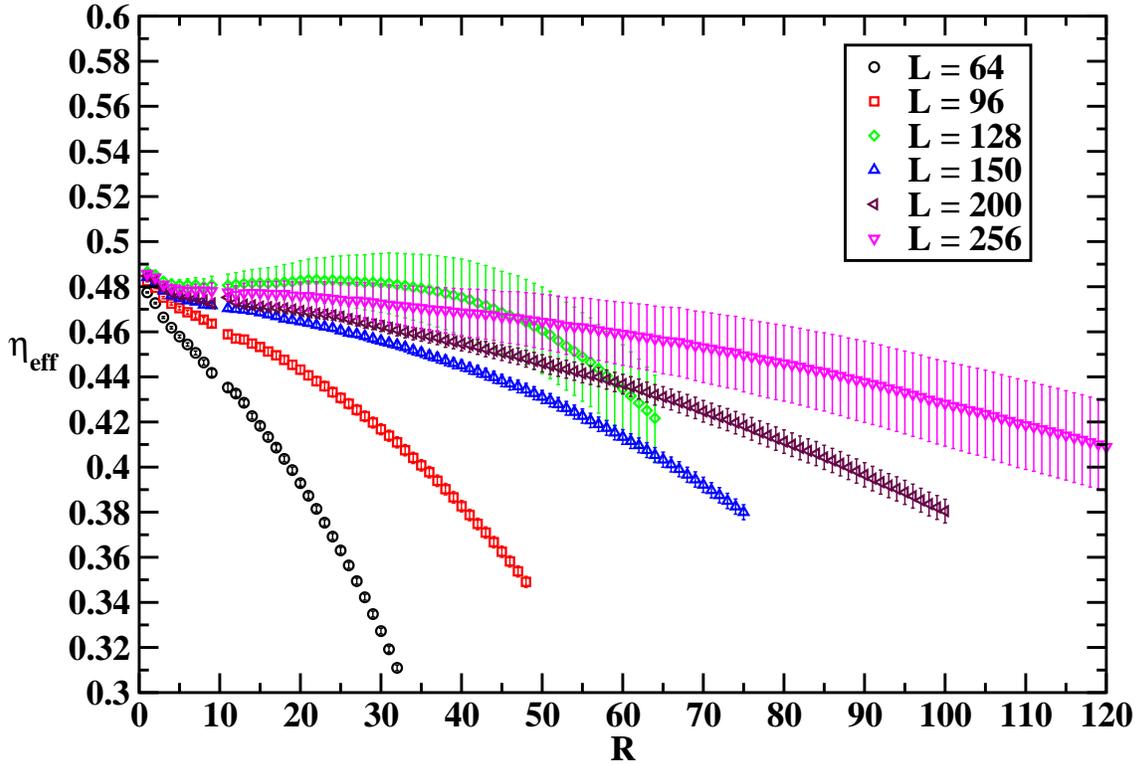}
\caption[]{$\eta_{\mbox{\scriptsize eff}}$ {\it vs} $R$ at $\beta_c$ on lattices with 
several values of $L$.}
\label{eta_eff_3.06}
\end{figure}

The analysis of the behaviour of $\eta_{\mbox{\scriptsize eff}}(R)$ has been
repeated setting $\beta$ at our estimated value for $\beta_c$, {\it i.e.} 
$\beta=3.06$, and increasing the spatial extent of the lattice. 
It turns out (see Fig.~\ref{eta_eff_3.06}) that a plateau develops at small 
distances when $L$ increases and that the extension of this plateau gets 
larger with $L$, consistently with the fact that finite volume effects are 
becoming less important. The plateau value of $\eta_{\mbox{\scriptsize eff}}$ 
can be estimated as $\eta_{\mbox{\scriptsize eff}}(R=6)$ on the 
256$^2\times 8$ lattice and is equal to 0.4782(25); it agrees with our 
previous determination of the index $\eta$.

\section{Discussions}

We have studied the critical behaviour of the $3d$ $U(1)$ LGT at finite 
temperature on isotropic lattice with the temporal extension $N_t=8$. 
The pseudo-critical coupling was determined through the peak in 
the susceptibility of the Polyakov loop; the infinite-volume critical coupling 
has then been computed assuming the scaling behaviour of the form (\ref{b_pc}),
the result being $\beta_c=3.06(11)$. The deconfinement phase is the phase 
where $\beta\geq\beta_c$. A thorough investigation of the deconfinement phase 
was performed in Ref.~\cite{chernodub}. However, all $\beta$-values used there 
are smaller than the infinite-volume critical coupling.
When the thermodynamic limit is approached the critical coupling increases so
that the numerical results of Ref.~\cite{chernodub} would refer rather to the
confinement phase of the infinite-volume theory.

We found also that the index $\eta$ turns out to be $\eta \approx 0.496$. 
This value is essentially larger than expected and requires some discussion. 
The easiest explanation would be to state that the spatial lattice size used 
($L\in [32-256]$) is still too small to exhibit the correct scaling behaviour, 
hence the wrong values for $\beta_c$ and $\eta$ follow. However, if one makes 
a plot of $\beta_{pc}(L)$ vs $L$, one can see, by looking at the trend of data,
that it is unlikely that $\beta_c$ is much larger than our estimate.
In fact, our fits with the scaling law (\ref{b_pc}) show that $\beta_c$ 
decreases when larger lattices are considered. Therefore, our result is most 
likely an overestimation. This implies that the true $\eta$ is likely even
larger than what we found. 

Moreover, neglecting logarithmic corrections to the scaling 
law~(\ref{eta_scale}) cannot have such a strong impact to decrease $\eta$ 
to half of the value we found. 

Let us give a simple argument why the index $\eta$ can be different from its 
$XY$ value. Consider the anisotropic lattice, in the limit of large 
$\beta_s=\infty$. Here the spatial plaquettes are frozen to unity and 
the ground state is a state where all spatial fields are pure gauge, i.e. 
$U_n(x) \ = \ V_x V^*_{x+e_n}  \ , \ n=1,2$. Under the change of variables 
$U_0(x)\to V_x U_0(x)V^*_{x+e_0}$, in the leading order of the large-$\beta_s$ 
expansion the partition function factorizes into the product of $N_t$ 
independent $2d$ $XY$ models. Since the Polyakov loop is the product of gauge 
fields in the temporal direction, the correlation function factorizes, too, 
and becomes a product of independent $XY$ correlations, i.e. 
\begin{equation}
\Gamma_{U(1)}(\beta_s=\infty , \beta_t) \ = \ \left [ \Gamma_{XY}(\beta_t) \right ]^{N_t} \ .
\label{PLcorr}
\end{equation}
Hence, for asymptotically large $R\gg 1$, we get
\begin{equation}
\Gamma_{U(1)}(\beta_s=\infty , \beta_t\geq\beta_t^{cr}) \ 
\asymp \ \left [ \frac{1}{R^{\eta_{XY}}} \right ]^{N_t} \ .
\label{PLcorr1}
\end{equation}
This leads to a simple relation
\begin{equation}
\eta (\beta_s=\infty,\beta_t^{cr}) \ = \ N_t \ \eta_{XY} \ .
\label{etabeta_sinfty}
\end{equation}

Some conclusions could now be drawn. The critical behaviour of the $3d$ $U(1)$ 
LGT in the limit $\beta_s\to\infty$ is also governed by the $2d$ $XY$ model. 
Nevertheless, the effective index $\eta$ appears to be $N_t$ times of its $XY$ 
value. 
Now, for $\beta_s=0$ we have $\eta (\beta_s=0,\beta_t^{cr}) \ = \  \eta_{XY}$. 
This relation and formula (\ref{etabeta_sinfty}) allow to conjecture that
\begin{equation}
\eta_{XY} \ \leq \ \eta (\beta_s,\beta_t^{cr}) \ \leq \ N_t \ \eta_{XY} \ .
\label{etabeta_s}
\end{equation}
$\beta_s=0$ corresponds to the lower limit while $\beta_s=\infty$ corresponds 
to the upper limit. In general, $\eta$ could interpolate between two limits 
with $\beta_s$. Whether this interpolation is monotonic or there exists 
critical value $\beta_s^{cr}$, such that 
$\eta (\beta_s\leq \beta_s^{cr},\beta_t^{cr}) \ = \ \eta_{XY}$ and $\eta$ 
changes monotonically above $\beta_s^{cr}$, cannot be answered with data we 
have and requires simulations on the anisotropic lattices. 
In the paper~\cite{rgu1} a renormalization group study of $3d$ $U(1)$ model 
at small $\beta_s$ will be presented and computations of the leading 
correction to the large $\beta_s$ behaviour will be given. The results of our 
computations support the scenario that the index $\eta$ depends on the ratio 
$\beta_s/\beta_t$. 
Recently, we have obtained the results of simulations for $N_t=2,4$ performed 
by A.~Bazavov~\cite{bazavov}. His results also point in the direction of our 
scenario.

Finally, it is worth mentioning that the factorization in the large $\beta_s$ 
limit does not affect the index $\nu$. It follows from its definition 
(\ref{PLlowt}) that in this limit $\nu=1/2$ as in the $XY$ model. 
We expect therefore that $\nu$ equals $1/2$ for all $\beta_s$ and is thus 
universal.

In view of our results it might be worth to perform numerical simulations 
for small but nonvanishing $\beta_s$ and for larger volumes.

\end{document}